\RequirePackage{fix-cm}
\documentclass[twocolumn]{svjour3}          
\smartqed  
\usepackage{graphicx}
\journalname{noname}
\begin{document}

\title{Actively tunable optical Yagi-Uda nanoantenna with bistable emission characteristics
}

\titlerunning{Actively tunable optical Yagi-Uda nanoantenna}        

\author{Ivan S. Maksymov         \and
		Andrey E. Miroshnichenko         \and
        Yuri S. Kivshar
}


\institute{Ivan S. Maksymov, Andrey E. Miroshnichenko and Yuri S. Kivshar \at
              Nonlinear Physics Centre, Research School of Physics and Engineering, Australian National University, Canberra ACT 0200, Australia \\
              Tel.: +61-2-6125-9074\\
              Fax: +61-2-6125-8588\\
              \email{mis124@physics.anu.edu.au}           
}

\date{Received: date / Accepted: date}

\maketitle

\begin{abstract}
We suggest and study theoretically a novel type of optical Yagi-Uda nanoantennas tunable via variation of the free-carrier density of a semiconductor disk placed in a gap of a metallic dipole feeding element. Unlike its narrowband all-metal counterparts, this nanoantenna exhibits a broadband unidirectional emission and demonstrates a bistable response in a preferential direction of the far-field zone, which opens up unique possibilities for ultrafast control of subwavelength light not attainable with dipole or bowtie architectures.
\keywords{Plasmonics \and Nanoantennas \and Yagi-Uda antennas \and Spectral tuning \and Bistability \and Ultrafast switches}
\end{abstract}

\section*{Introduction}
\label{intro}
Owing to recent advances in nanotechnology, plasmonic nanoantennas have become a subject of considerable theoretical
and experimental interest \cite{gia11_review,nov11_review}. Plasmonic resonances in nanoantennas allow breaking through the fundamental diffraction limit \cite{gra10}, opening up novel opportunities for controlling light--matter interactions within subwavelength volumes. Several potential applications of nanoantennas have been considered in topics such as spectroscopy and high-resolution near-field microscopy \cite{nov11_review}, subwavelength light confinement and enhancement \cite{sto04}, photovoltaics \cite{atw10}, sensing \cite{liu11}, molecular response enhancement \cite{ang10}, non-classical light emission \cite{mak10}, and communication \cite{alu10_prl}. 

In many of these applications, controlling and modifying the far-field of a nanoantenna is an important issue that is particularly interesting for obtaining directional beaming effects, which have been demonstrated e.g. with Yagi-Uda architectures \cite{koe09,cur10,kos10,dre11,dor11}. However, an efficient nanoantenna must not only have a large local field enhancement and a high directivity \cite{mak11_1} but also be wavelength tunable over a wide spectral range \cite{mir11} because it allows a smaller nanoantenna to behave as a larger nanoantenna or as an array of nanoantennas \cite{dre11}, both saving space and improving performance. 

Consequently, a large and growing body of research investigates tunable nanoantennas \cite{far05,hua10,alu08,lar10,abb11,she11,mak11_oe,ala11}. Many novel control mechanisms try to exploit the concept of metamaterial-based \cite{zio08} and non-foster impedance matching circuits \cite{sus09}, where one of the possible ways for achieving spectral tuning consists in the use of tunable nanocapacitors and/or nanoinductors \cite{eng05,eng_science}. Other approaches may rely on vanadium oxide tunable metamaterials \cite{seo10}, mechanically reconfigurable photonic metamaterials \cite{ou11} or metamaterials hybridized with carbon nanotubes \cite{nik10}. Large spectral tunability can also be obtained using electrically controlled liquid crystals \cite{kos05,ber09,ang11}, but a very slow response of liquid crystals is not suitable for many application and, in general, a solid-state implementation is more suitable for on-chip integration of nanoantennas.

However, the spectral tuning of optical Yagi-Uda nanoantennas has not so far been demonstrated because their capability to tune to multiple operating frequencies is compromised by their ability to receive and transmit light in a preferential direction (their operating bandwidth is limited to just a few percents around the designed resonance frequency). Tunable Yagi-Uda nanoantennas could have technological applications in building broadband optical wireless communication systems,  advanced nano-sensor systems, high performance solar cells as well as wavelength tunable single photon sources and detectors. 

In this Letter, we suggest to exploit the free carrier nonlinearity of semiconductors for a dynamical tuning of the operating wavelength of a plasmonic unidirectional Yagi-Uda nanoantenna consisting of silver nanorods used for the feeding element, reflector and directors. We modify the feeding element as compared with previous designs \cite{cur10} and consider a semiconductor nano-disk squeezed by two identical nanorods. The illumination of the feeding element with a laser beam alters the conductivity of the nano-disk and enables a monotonic tuning of the operating wavelength of the nanoantenna in a very wide spectral range as compared with that of conventionally designed Yagi-Uda nanoantennas \cite{cur10,kos10,dre11,dor11}. It also gives rise to a bistable response of the nanoantenna in the far-field zone. Along with the unidirectional emission properties of the nanoantenna, the  bistability opens up novel opportunities for ultra-fast switching, optical limiting, logic gating, modulation and amplification of light pulses \cite{gib85}.

\section*{Design and Simulation Model}
\label{sec:1}
The mainstream of the design and interpretation of the results are performed using CST Microwave Studio software implementing a Finite Integration Technique. Figure $1$ shows a plasmonic Yagi-Uda nanoantenna consisting of silver nanorods used for reflector, feeding element and directors. The length of the nanorods is designed relying on the effective wavelength rescaling principle \cite{nov11_review}, which takes into account the volume of the nanorods and absorption losses in silver. According to this approach, the resulting reduced effective wavelength $\lambda_{eff}$ seen by the nanoantenna is related to the incident wavelength $\lambda$ by a simple relation $\lambda_{eff}=n_{1}+n_{2}\frac{\lambda}{\lambda_{p}}$, where $n_1$ and $n_2$ are geometric constants and $\lambda_p$ is the plasma wavelength.

\begin{figure}
\includegraphics[width=8cm]{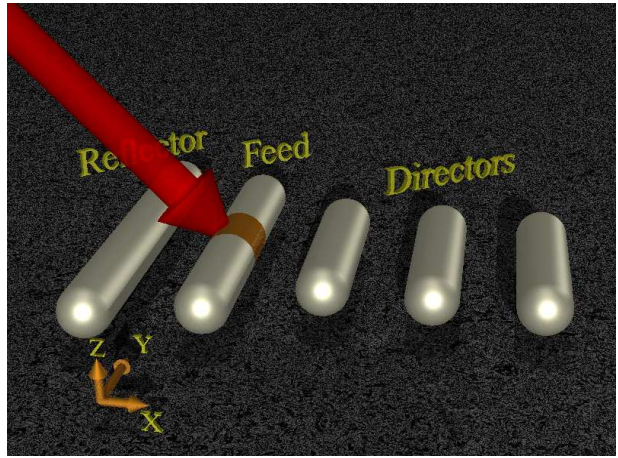}
\caption{A tunable plasmonic Yagi-Uda nanoantenna consisting of silver nanorods used for reflector, feeding element and directors. The yellow area of the feeding elements corresponds to the semiconductor nano-disk used as loading. The nanoantenna is surrounded by air. The semi-transparent red arrow schematically shows the direction of the incident plane wave. }
\label{fig:1}
\end{figure}

We optimize the nanoantenna performance for a central wavelength to be $\approx1\mu$m. We choose the radii of the nanorods and those of their outer rounded edges as $r=25$ nm, and the spacing between all elements as $w=30$ nm \cite{mak11_1}. The feeding element consists of two nanorods with non-rounded inner edges, separated by a semiconductor nano-disk. The total length of the feeding elements including the nano-disk is $L=390$ nm. The lengths of the reflector and directors are chosen as $1.125L$ and $0.75L$, respectively. We assume that the nano-disk of the feeding element is made of amorphous silicon (a-Si) and that its width constitutes $50$ nm. The nanoantenna is surrounded by air because this provides the simplest model to which additional elements of any practical design, such as e.g. a substrate, can be added.

A change in the dielectric permittivity of the nano-disk loading of the feeding element caused by an increase in the free carrier density is modelled using a Drude model based on experimental values $\epsilon_{exp}(\omega)$ of a-Si \cite{pal91} as
\begin{eqnarray}
\epsilon(\omega)=\epsilon_{exp}(\omega)-(\frac{\omega_{pl}}{\omega})^2\frac{1}{1+i\frac{1}{\omega\tau_{D}}}
\label{eq:one},
\end{eqnarray}
where $\omega_{pl}=\sqrt{Ne^2/\epsilon_{0}m_{opt}^{*}m_{e}}$ denotes the plasma frequency, with $N$ the free carrier concentration. The rest of the material parameters and a relevant discussion of their choice can be found in Ref. \cite{lar10}.

\section*{Results and Discussion}
\label{sec:2}
To start with, we investigate the performance of the isolated feeding element. We observe that a strong modification of the free carrier density profile leads to the transition of the feeding element from a primarily capacitive (dielectric) to a primarily conductive (metallic) mode. In a perfect agreement with Ref. \cite{lar10} we observe this transition when $Re \epsilon(\omega)=0$.

As a next step, we investigate near- and far-field zone characteristics of the Yagi-Uda nanoantenna. We consider the free carrier density in between $0$ and $3\cdot10^{21}$ cm$^{-3}$, that is in the range where the feeding elements switches from the capacitive mode to the conductive one. The nanoantenna is illuminated with a linearly polarized plane wave (the electric field is orientated along the \textit{y}-coordinate) incident from the rear end of the nanoantenna under the angle of $45$ degrees as shown in Fig. $1$.

The top panel of Fig. $2$ shows the power emitted by the nanoantenna in the maximum emission direction designed to align with the \textit{x}-axis. The bottom curve was calculated for the free carrier density of $0$ cm$^{-3}$ whereas the others  were calculated for the free carrier densities gradually increasing up to $3\cdot10^{21}$ cm$^{-3}$. In all curves we observe three maxima corresponding to different operating resonance wavelengths. Since the nanoantenna is designed to perform around the wavelength of $1\mu$m, we first investigate the resonances occurring in this wave range. We notice that an increase in the free carrier density produces a monotonic decrease in the operating wavelengths from $1.07\mu$m (denoted as A in Fig. $2$) to $0.84\mu$m (denoted as A') also accompanied by a weak decrease in the power emitted by the nanoantenna.

\begin{figure*}
\includegraphics[width=12cm]{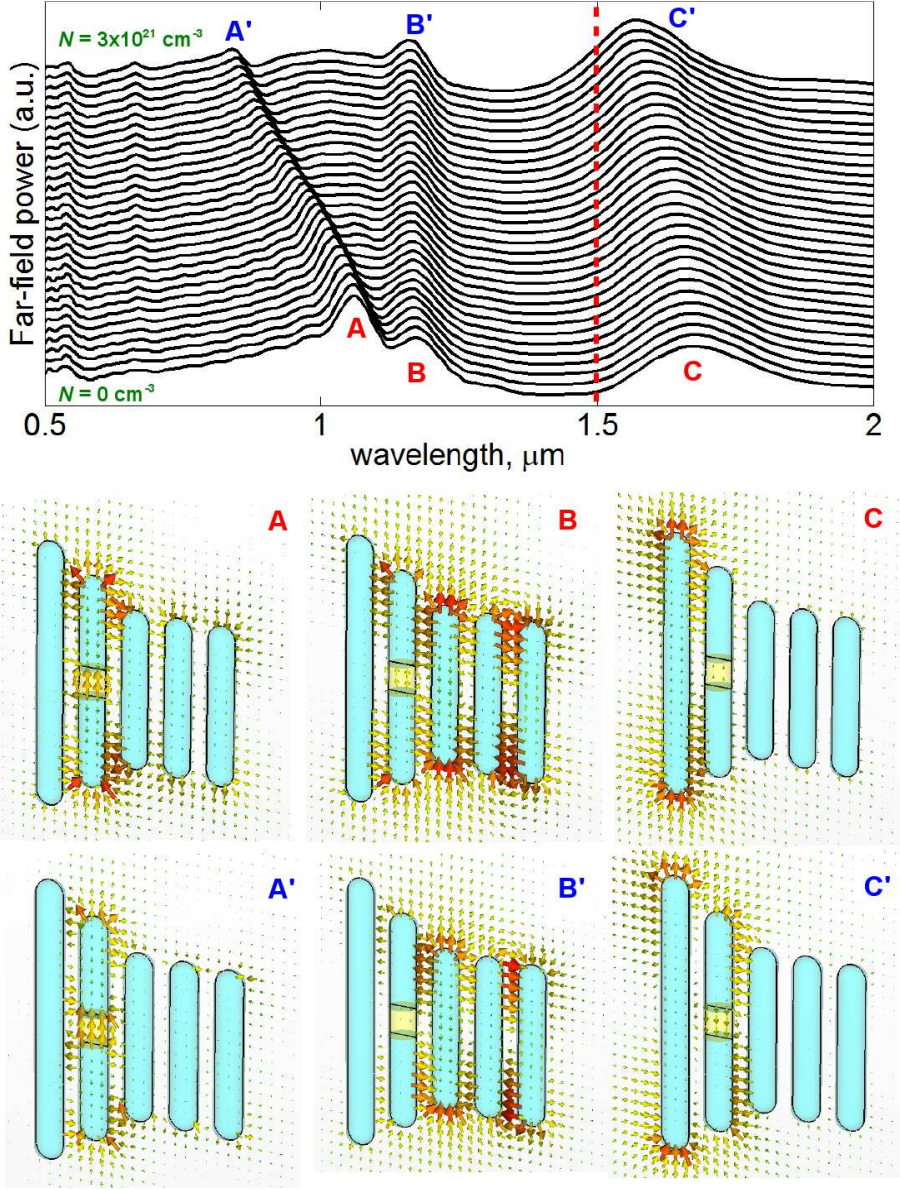}
\caption{(Top) Far-field power spectra of the nanoantenna in the maximum emission direction as a function of the wavelength for different free carrier densities from $0$ cm$^{-3}$ to $3\cdot10^{21}$ cm$^{-3}$ (from bottom curve and
up). Red dashed line indicates the operating wavelength of a Yagi-Uda nanoantenna with the same dimensions but equipped with an all-metal feeding element. (Bottom) Stationary $|E|$ electric field distributions in the near-field zone of the nanoantenna corresponding to the resonance peaks denoted by the capital letters in the top panel.}
\label{fig:2}
\end{figure*}

In order to confirm that in this wave range the feeding element is in its resonance as should be in the case of a Yagi-Uda antenna \cite{dor11}, in the bottom panel of Fig. $2$ we plot stationary $|E|$ electric field distributions in the near-field zone of the nanoantenna. As shown in sub-panels A and A' (here A corresponds to $0$ cm$^{-3}$ and A' to $3\cdot10^{21}$ cm$^{-3}$), the feeding element is at its maximum field strength brighter than all the other elements.

These results demonstrate that the resonance wavelength of the A--A' resonances is gradually blue-shifted, which means that the nanoantenna can be gradually tuned in a $230$ nm wave range by controlling the loading of the feeding element. The spectral tuning can be understood using an intuitive flat-plate capacitor model where the inner faces of the nanorods of the feeding element play the role of the capacitor's plates. In a capacitor, the electric current leads the polarization charges by $\pi/2$ that means that the charge reduces to zero when the current reaches its maximum. In the framework of our model, an increase in the free carrier density gives rise to a current, which reduces the charges as well as reduces the effective volume of the capacitive loading. The feeding element with a smaller loading starts to support another mode at a shorter wavelength and with a slightly lower amplitude \cite{lar10}. It is obvious that the free carrier densities of more than $3\cdot10^{21}$ cm$^{-3}$ will maximize the current and the mode supported by the feeding element will die out. For this reason in what follows we discuss the free carrier densities of up to $3\cdot10^{21}$ cm$^{-3}$ only.

Importantly for our further analysis, in analogy to a flat-plate capacitor the electric field between the inner faces of the nanorods of the feeding elements is nearly uniform, as shown by the lines of field in Fig. $2$. This finding will significantly simplify simulations of the nanoantenna.

The analysis of resonances denoted by B--B' (see the top panel of Fig. $2$) does not reveal a significant change in the resonance wavelengths because the electric field is mainly concentrated around the directors of the nanoantenna and does not penetrate into the semiconductor, as shown in the bottom panel of Fig. $2$.

The C--C' resonances are more sensitive to a variation of the conductivity of the nano-disk due to the fact that at $0$ cm$^{-3}$ the electric field mainly concentrates around the reflector but spreads over both the reflector and the feeding element at $3\cdot10^{21}$ cm$^{-3}$. The wavelength of C--C' resonances is blue-shifted towards the operating wavelength of an all-metal Yagi-Uda with the same geometry (red dashed line in the top panel of Fig. $2$).

Since both B--B' and C--C' resonances the nanoantenna do not exhibit a correct Yagi-Uda behavior manifesting itself by a pronounced resonance of the feeding element \cite{dor11}, in what follows we focus ourselves on A--A' resonances only.

It is worth to notice a highly desirable option of the nanoantenna excitation with a broadband point-like emitter (e.g. a fluorescent molecule) placed near one of the edges of the feeding element \cite{li09,mir11}. It allows a realization of a pump-probe operation scheme, where a pump laser is used to control the coupling of the emitter to the nanoantenna at different wavelengths. Moreover, according to the principle of reciprocity the far-field characteristics of the nanoantennas in the emission regime are similar to those in the reception one \cite{nov11_review}, and, therefore, the nanoantenna can be employed as a tunable nano-receiver.

In active semiconductor nanophotonic devices, such as e.g. all-optical switches based on photonic crystal nano-cavities (see e.g. Ref. \cite{bel08}) and semiconductor antennas for THz radiation \cite{ber10}, one usually needs free carrier densities of up to $10^{19}$ cm$^{-3}$. As we can see in Fig. $2$, in order to tune the response of the Yagi-Uda nanoantenna by $\approx200$ nm one needs to increase the free carrier density by two orders of magnitude as compared with that for the aforementioned nanophotonic devices. Hereafter, we demonstrate that this increase can be achieved at experimentally attainable optical intensities owing to a local field enhancement in the nano-disk loading of the feeding element.

For our further analysis, we consider that the free carrier density $N$ in the semiconductor obeys the rate equation \cite{sze69}
\begin{eqnarray}
\frac{\partial N(t)}{\partial t}=-\frac{N(t)}{\tau_{c}} + \frac{c^2 {\epsilon_0}^2 {n_0}^2 \beta_{TPA}}{8 \hbar \omega_0}|E(t)|^4
\label{eq:two},
\end{eqnarray}
where $\tau_{c}$ is the free carrier lifetime, $E(t)$ is the amplitude of the electric field, $\omega_0$ is the angular frequency of the excitation plane wave and $\beta_{TPA}$ is the two-photon absorption coefficient. We take $\tau_{c}=1$ ns and $\beta_{TPA}=120$ cm/GW \cite{lar10}. Owing to the uniformity of the electric field in the nano-disk loading (see Fig. $2$) in the simulations it is safe to assume that the free carrier distribution in the nanodisk is also uniform. It makes it possible to find a self-consistent solution to the nonlinear problem of the free carrier dynamics in the semiconductor using to the following numerical procedure.

First, from the steady-state rate equation we find values of $|E|$ for $N=0\ldots3\cdot10^{21}$ cm$^{-3}$. Secondly, we use CST Studio where we fix the wavelength and the amplitude of the incident plane wave, which is a constant in all numerical experiments, and carry out simulations in order to find steady-state values of the electric field $E_{d}$ in the nano-disk. Then, we calculate the ratio between the electric field amplitude of the incident wave and the electric field induced by this wave in the nano-disk as $\alpha =  \frac{|E|}{|E_{d}|}$. Finally, using the relation $E_{inc}=\alpha E_{0}$ we derive the real amplitudes of the plane wave that should be applied to the feeding element in order to induce the free carrier densities of up to $3\cdot10^{21}$ cm$^{-3}$. We have neglected here and in the following the possible nonlinear effects in metal, which are considered negligible compared to the nonlinearities in a-Si. It is also worth to note that this approach automatically takes into consideration an energy shift between near- and far-field zone peak powers \cite{mir10,zul10}.

Figure $3$(a) shows the steady-state dependencies of the free carrier concentration on the optical intensity obtained using the suggested numerical procedure. The excitation of the nanoantenna in the wave range between $0.95 \mu$m and $1.05\mu$m covering the A--A' resonances (Fig. $2$) leads to a bistable response between the optical intensity and the free carrier density in the nano-disk. As can be seen in Fig. $3$(b), which shows absolute values of the electric field in the nano-disk loading as a function of the free carrier density, the observed multi-stable behavior is a result of an enhancement of the local field in the nano-disk squeezed between two silver nanorods. The maximum field enhancement takes place at the shortest of the considered operating wavelengths and gradually decreases with an increase in the wavelength.

\begin{figure}
\includegraphics[width=5cm]{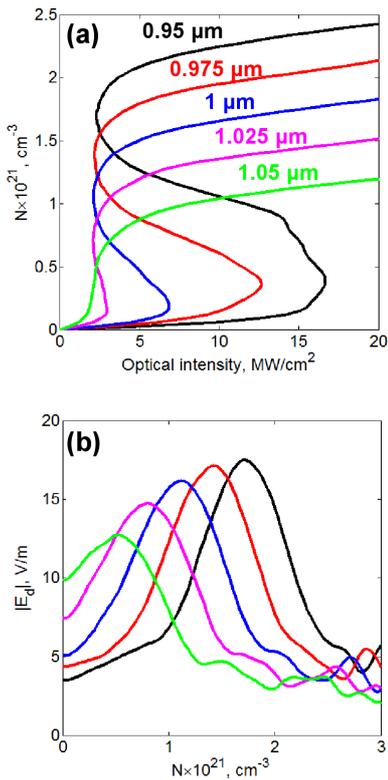}
\caption{(a) Free carrier density in the nano-disk loading of the feeding elements as a function of the optical intensity at different operating wavelengths. (b) Absolute value of the electric field in the nano-disk loading as a function of the free carrier density at different operating wavelengths.}
\label{fig:3}
\end{figure}

It is worth mentioning here that we also investigated the impact of the background material on the relation between the free carrier concentration and the optical intensity. We found that the presence of a background does not significantly change the performance of the nanoantenna apart from red-shifting its operating wavelengths.

In order to gain more insight into the far-field characteristics of the nanoantenna, in Figs. $4$(a-e) we plot its far-field power angular diagrams calculated for different operating wavelengths. It is important for the spectral tuning that in all regimes the nanoantenna performs as an unidirectional emitter with a high front-to-back ratio and nearly constant beam-width of $\approx80^o$.  Moreover, by plotting the power emitted by the nanoantenna in the maximum emission direction as a function of the optical intensity [see Figs. $4$(f-j)], we observe the formation of closed bistability loops at different operating wavelengths.

The formation of closed loops was  observed earlier in photonic systems exhibiting nonlinear Fano-Feshbach resonances resulting from the interaction between two Fano resonances located very close to each other \cite{mir09}. The appearance of the closed loops in the far-field characteristics of the Yagi-Uda nanoantenna can be also understood using a simplest capacitor model. Here, a decrease in the far-field power with an increase in the optical intensity is due to an inevitable reduction of the effective volume of the capacitive loading of the feeding element.

\begin{figure*}
\includegraphics[width=16cm]{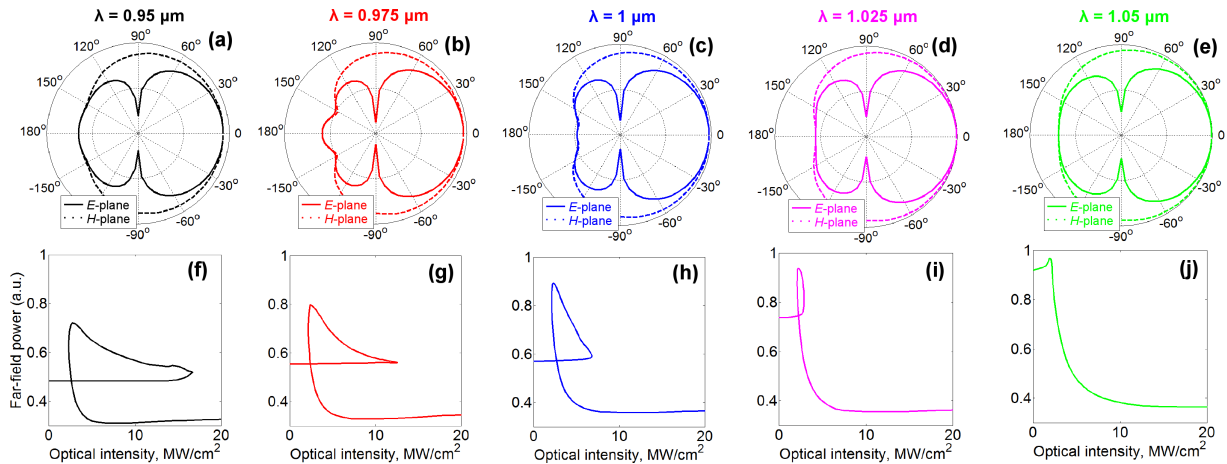}
\caption{ (a-e) Far-field power angular diagram of the nanoantenna in the E-plane (solid curves) and H-plane (dashed curves) at operating wavelengths (free carrier densities) of $0.95\mu$m ($1.9\cdot10^{21}$ cm$^{-3}$), $0.975\mu$m ($1.5\cdot10^{21}$ cm$^{-3}$), $1\mu$m ($1.2\cdot10^{21}$ cm$^{-3}$), $1.025\mu$m ($0.8\cdot10^{21}$ cm$^{-3}$) and $1.05\mu$m ($0.55\cdot10^{21}$ cm$^{-3}$). A dB scale is used to emphasize the difference in the backward lobes. (f-j) Power emitted by the nanoantenna in the maximum emission direction as a function of the optical intensity.}
\label{fig:4}
\end{figure*}

The closed bistable loops observed in Figs. $4$(f-i) may have appealing applications in realizing ultra-fast all-optical switching devices at the nanoscale since the nanoantenna exhibits two different stable states for the same applied optical intensity. One relevant aspect to underline here consists in the dependence of the total hysteresis area on the operating wavelength. For certain operating wavelength the hysteresis loop may not appear, as shown in Fig. $4$(j) for $1.05\mu$m (green curve). Such steep properties of the nonlinear response suggest the use of the Yagi-Uda architecture shown in Fig. $1$ as a nonlinear optical device operating as a logical cell, an optical limiter or a signal amplitude modulator \cite{gib85}, also exhibiting emission properties not attainable with dipole plasmonic nanoantennas \cite{che10,zho10,seo10}.

Finally, we calculate the optical intensity that maintains the nanoantenna in the appropriate operating regime. By fixing the operating wavelength at $1.05\mu$m and choosing a bias point in the steep part of the corresponding curve in Fig. $4$(j), for the excitation with a 25-ps-long pump laser beam focused to a $1$ micron squared spot, we obtain the trigger energy of $\approx1$ pJ. Bistability-based operation of the nanoantenna at shorter wavelengths would require higher pump energies of up to $\approx5$ pJ. These values are achievable in practice and are consistent with the requirements for the ideal bistable optical device \cite{gib85}.

\section*{Conclusions}
In conclusion, we have suggested a simple way to tune dynamically a plasmonic Yagi-Uda nanoantenna and emit light in a wide spectral range. This capability is hardly achievable with conventionally designed Yagi-Uda nanoantennas whose performance is strictly optimized to just a few percent around the design frequency, and it cannot be extended significantly without a penalty. We have shown the capability of the optical Yagi-Uda nanoantennas to perform as a bistable optical device offering new degrees of freedom in controlling the far-field emission. As such, Yagi-Uda nanoantennas can be used for ultra-fast switching, mixing, frequency conversion, modulation and other kinds of all-optical light control and manipulation at the nanoscale. We have shown that the optical energy required to switch the nanoantenna to an unstable state is achievable in experiments.

\begin{acknowledgements}
This work was supported by the Australian Research Council. The authors confirm many valuable discussions with their colleagues from the Nonlinear Physics Centre and Metamaterial Meeting Group at the Australian National University.
\end{acknowledgements}



\begin{thebibliography}{00}

\bibitem{gia11_review} Giannini V, Fern{\'{a}}ndez-Dom{\'{i}}nguez AI, Heck SC, Maier SA (2011) Plasmonic nanoantennas: Fundamentals and their use in controlling the radiative properties of nanoemitters. Chem Rev 111:3888--3912

\bibitem{nov11_review} Novotny L, van Hulst NF (2011) Antennas for light. Nat Photonics 5:83--90 

\bibitem{gra10} Gramotnev DK, Bozhevolnyi SI (2010) Plasmonics beyond the diffraction limit. Nat Photonics 4:83--91 

\bibitem{sto04} Stockman MI (2004) Nanofocusing of optical energy in tapered plasmonic waveguides. Phys Rev Lett 93:137404

\bibitem{atw10} Atwater HA, Polman A (2010) Plasmonics for improved photovoltaic devices. Nat Mater 9:205--213

\bibitem{liu11} Liu N, Tang ML, Hentschel M, Giessen H, Alivisatos AP (2011) Nanoantenna-enhanced gas sensing in a single tailored nanofocus. Nat Mater 10:631--636

\bibitem{ang10} De Angelis F, Das G, Candeloro P, Patrini M, Galli M, Bek A, Lazzarino M, Maksymov IS, Liberale C, Andreani LC, di Fabrizio E (2010) Nanoscale chemical mapping using three-dimensional adiabatic compression of surface plasmon polaritons. Nat Nanotechnol 5:67--72

\bibitem{mak10} Maksymov IS, Besbes M, Hugonin JP, Yang J, Beveratos A, Sagnes I, Robert-Philip I, Lalanne P (2010) Metal-coated nanocylinder cavity for broadband nonclassical light emission. Phys Rev Lett 105:180502

\bibitem{alu10_prl} Al{\'{u}} A, Engheta N (2010) Wireless at the nanoscale: {O}ptical interconnects using matched nanoantennas. Phys Rev Lett 104:213902

\bibitem{koe09}  Koenderink AF (2009) Plasmon nanoparticle array waveguides for single photon and single plasmon sources. Nano Lett 9:4228--4233

\bibitem{cur10} Curto AG, Volpe G, Taminiau TH, Kreuzer MP, Quidant R, van Hulst N F (2010) Unidirectional emission of a quantum dot coupled to a nanoantenna. Science 329:930--933

\bibitem{kos10} Kosako T, Kadoya Y, Hofmann HF (2010) Directional control of light by a nano-optical {Y}agi–{U}da antenna. Nat Photonics 4:312--315

\bibitem{dre11} Dregely D, Taubert R, Dorfm{\"{u}}ller J, Vogelgesang R, Kern K, Giessen H (2011) 3{D} optical {Y}agi-{U}da nanoantenna array. Nat Commun 2:267

\bibitem{dor11} Dorfm{\"{u}}ller J, Dregely D, Esslinger M, Khunsin W, Vogelgesang R, Kern K, Giessen H (2011) Near-field dynamics of optical Yagi-Uda nanoantennas. Nano Lett 11:2819--2824

\bibitem{mak11_1} Maksymov IS, Davoyan AR, Kivshar YuS (2011) Enhanced emission and light control with tapered plasmonic nanoantennas. Appl Phys Lett 99:083304

\bibitem{mir11} Miroshnichenko AE, Maksymov IS, Davoyan AR, Simovski C, Belov P, Kivshar YuS (2011) An arrayed nanoantenna for broadband light emission and detection. Phys Status Solidi--Rapid Res Lett 5:347--349

\bibitem{far05} Farahani JN, Pohl DW, Eisler H--J, Hecht B (2005) Single quantum dot coupled to a scanning optical antenna: A tunable superemitter. Phys Rev Lett 95:017402

\bibitem{hua10} Huang F, Baumberg JJ (2010) Actively tuned plasmons on elastometrically driven Au nanoparticle dimers. Nano Lett 10:1787--1792

\bibitem{alu08} Al{\'{u}} A, Engheta N (2008) Input impedance, nanocircuit loading, and radiation tuning of optical nanoantennas. Phys Rev Lett 101:043901

\bibitem{lar10} Large N, Abb M, Aizpurua J, Muskens OL (2010) Photoconductively loaded plasmonic nanoantenna as building block for ultracompact optical switches. Nano Lett 10:1741--1746

\bibitem{abb11} Abb M, Albella P, Aizpurua J, Muskens OL (2011) All-optical control of a single plasmonic nanoantenna -- {ITO} hybrid. Nano Lett 11:2457--2463

\bibitem{she11} Shegai T, Chen S, Milkovi\'{c} V, Zengin G, Johansson P, K\"{a}ll M (2011) A bimetallic nanoantenna for directional colour routing. Nat Commun 2:481

\bibitem{mak11_oe} Maksymov IS, Miroshnichenko AE (2011) Active control over nanofocusing with nanorod plasmonic antennas. Opt Express 19:5888--5894

\bibitem{ala11}  Alaverdyan Y, Vamivakas N, Barnes J, Lebouteiller C, Hare J, Atat\"{u}re M (2011) Spectral tunability of a plasmonic antenna with a dielectric nanocrystal. Opt Express 19:18175--18181

\bibitem{zio08} Ziolkowski RW, Erentok A (2008) Metamaterial-based efficient electrically small antennas. IEEE Trans Antennas Propag 54:2113--2130

\bibitem{sus09} Sussman-Fort SE, Rudish RM (2009) Non-foster impedance matching of electrically-small antennas. IEEE Trans Antennas Propag 57:2230--2241

\bibitem{eng05} Engheta N, Salandrino A, Al{\'{u}} A (2005) Circuit elements at optical frequencies: {N}anoinductors, nanocapacitors, and nanoresistors. Phys Rev Lett 95:095504

\bibitem{eng_science} Engheta N (2007) Circuits with light at nanoscales: {O}ptical nanocircuits inspired by metamaterials.  Science 317:1698--1702

\bibitem{seo10} Seo M, Kyoung J, Park H, Koo S, Kim H-S, Bernien H, Kim BJ, Choe JH, Ahn YH, Kim H-T, Park N, Park Q-H, Ahn K, Kim D-S (2010) Active terahertz nanoantennas based on {VO}2 phase transition. Nano Lett 10:2064--2068

\bibitem{ou11} Ou JY, Plum E, Jiang L, Zheludev NI (2011) Reconfigurable photonic metamaterials. Nano Lett 11:2142--2144

\bibitem{nik10} Nikolaenko AE, de Angelis F, Boden SA, Papasimakis N, Ashburn P, di Fabrizio E, Zheludev NI (2010) Carbon nanotubes in a photonic metamaterial. Phys Rev Lett 104:153902

\bibitem{kos05} Kossyrev PA, Yin A, Cloutier SG, Cardimona DA, Huang D, Asling PM, Xu JM (2005) Electric field tuning of plasmonic response of nanodot array in liquid crystal matrix. Nano Lett 5:1978--1981

\bibitem{ber09} Berthelot J, Bouhelier A, Huang C, Margueritat J, Colas-des-Francs G, Finot E, Weeber J--C, Dereux A, Kostcheev S, Ahrach H. Ibn El, Baudrion A-L, Plain J, Bachelot R, Royer P, Wiederrecht G P (2009) Tuning of an optical dimer nanoantenna by electrically controlling its load impedance. Nano Lett 9:3914--3921

\bibitem{ang11} De Angelis C, Locatelli A, Modotto D, Boscolo S, Midrio M, Capobianco A--D (2011) Frequency addressing of nano-objects by electrical tuning of optical antennas. J Opt Soc Am B 27:997--1001

\bibitem{gib85} Gibbs HM (1985) Optical bistability: {C}ontrolling light with light. Academic Press, Orlando

\bibitem{pal91} Palik ED (1985) Handbook of optical constants of solids. Academic Press, New York

\bibitem{li09} Li J, Salandrino A, Engheta N (2009) Optical spectrometer at the nanoscale using optical {Y}agi-{U}da nanoantennas. Phys Rev B 79:195104

\bibitem{bel08} Belotti M, Galisteo--L\'{o}pez JF, de Angelis S, Galli M, Maksymov IS, Andreani LC, Peyrade D, Chen Y (2008) All-optical switching in 2{D} silicon photonic crystals with low loss waveguides and optical cavities. Opt Express  16:11624--11636

\bibitem{ber10}  Berrier A, Ulbricht R, Bonn M, G\'{o}mez--Rivas J (2010) Ultrafast active control of localized surface plasmon resonances in silicon bowtie antenna. Opt Express 18:23226--23235

\bibitem{sze69} Sze SM (1969) Physics of semiconductor devices. John Wiley and Sons, New York

\bibitem{mir10} Miroshnichenko AE (2010) Off-resonance field enhancement by spherical nanoshells. Phys Rev A 81:053818

\bibitem{zul10} Zuloaga J, Nordlander P (2010) On the energy shift between near-field and far-field peak intensities in localized plasmon systems. Nano Lett 11:1280--1283

\bibitem{mir09} Miroshnichenko AE (2009) Nonlinear {F}ano-{F}eshbach resonances. Phys Rev E 79:026611

\bibitem{che10} Chen P--Y, Al{\'{u}} A (2010) Optical nanoantenna arrays loaded with nonlinear materials. Phys Rev B 82:235405

\bibitem{zho10} Zhou F, Liu Y, Li Z--Y, Xia Y (2010) Analytical model for optical bistability in nonlinear metal nano-antennae involving {K}err materials. Opt Express 13:13337--13344

\end{thebibliography}
\end{document}